\documentstyle[prb,eqsecnum,aps]{revtex}
\begin{document}
\title{A New Approach to the Spin $\frac{1}{2}$ Quantum Heisenberg Model}
\author{K.-H. M\"utter\cite{muetter}}
\address{Physics Department, University of Wuppertal\\
D-5600 Wuppertal 1, Germany\\[0.5cm]}

\date{WU B 93-11, Mai 93, cond-mat/9303039}

\maketitle

\begin{abstract}
An evolution equation for the expectation values of the Boltzmann factor
between monomer valence
bond states is derived. It contains the whole information on the
thermodynamical and magnetic
properties of the spin $\frac{1}{2}$ quantum Heisenberg model.
A new approximation scheme is
proposed and studied in the thermodynamical limit.
\end{abstract}

\section{Introduction}
In 1938 Hulth\'en \cite{hul} determined eigenvalues and eigenstates of the
spin $\frac{1}{2}$ quantum Heisenberg model on small clusters with
$V\leq 10$ sites. At that time, computers were not yet at his disposal.
He therefore exploited the symmetries of the model-including the
conservation of total spin $\vec{S}$- in order to reduce the
dimensions of the block Hamiltonians as much as possible. With the
advent of powerful computers exact diagonalizations on larger
clusters up to $V=36$ sites \cite{sch} and quantum Monte Carlo simulations
on much larger systems \cite{din} could be performed. Many of these
computations
were devoted to the determination of the ground-state properties of
the antiferromagnetic model: the energy and staggered magnetization
\cite{wie},
spin-spin correlators \cite{bor} etc. Since the ground-state is known to have
spin $0$ \cite{lie} it is sufficient to diagonalize the Hamiltonian in this
sector. Conservation of total spin, however, is not used in most of these
computations. An exception is Andersons resonating valence bond (RVB) state
\cite{and}. It seems that Hulth\`en's construction of spin $0$ states by means
of valence bonds is not so suited for numerical calculations on large
systems for the following reasons:
\begin{itemize}
\item The set of valence bond states is overcomplete; there are
    linear dependences among them.
\item Valence bond states are not orthogonal to each other.
\end{itemize}
There have been made some attempts to overcome these deficiencies \cite{sai}.
E.g. one can try to restrict the number of `allowed' valence bond
configurations in order to avoid over completeness. Such a procedure
was successful for $d=1\;$ \cite{sai}. For $d>1$, however, it looks very
unpleasant: The apparent symmetries of the system -translations, rotations,
reflections are hidden in complicated relations.\\
In this paper I want to draw the attention to some astonishing properties
of the projection operators on valence bond states:
\begin{itemize}
\item Completeness
\item Spectral like representations of the Hamiltonian and other
      operators.
\item Anticommutation relations with the Hamiltonian.
\end{itemize}
The properties are discussed in section 2 and extended to projection
operators on spin $S$ states in section 3. In section 4 these properties
are shown to describe the evolution of the expectation values of the
Boltzmann factor between monomer valence bond states.
In section 5 an
approximation scheme for the evaluation of the evolution equation
expansion is proposed: the monomer dimer cluster expansion.
The thermodynamical
limit of the lowest order approximation is investigated in section 6.
\section{Valence Bond Projection Operators (VBPO)}
We consider a lattice with an even number $V=2v$ of sites. Spin operators
$\vec{\sigma}(x)$ live at each site $x$. The operator
\begin{equation}
(x,y)=\frac{1}{4} (1-\vec{\sigma}(x) \vec{\sigma}(y))
\end{equation}
projects the two spin state at sites $x$ and $y$ onto its
valence bond (i.e. spin $0$ ) part:
\begin{equation}
[x,y]=\frac{1}{\sqrt{2}}(\chi_+(x) \chi_-(y)-\chi_-(x) \chi_+(y)).
\end{equation}
Each partition of the lattice into pairs $[x_j,y_j]$ of sites defines
a valence bond configuration (VBC) $K$. We associate to each configuration
$K$ a state:
\begin{equation}
|K\rangle=[x_1,y_1][x_2,y_2]...[x_v,y_v],
\end{equation}
with total spin $0$. These valence bond states yield an overcomplete
non orthogonal basis in the total spin $0$ sector.
In the following we are concerned with the valence bond projection
operators (VBPO):
\begin{equation}
P(K)=|K\rangle\langle K|=(x_1,y_1)(x_2,y_2)...(x_v,y_v).
\end{equation}
The VBPO have the following properties:
\begin{equation}
{\rm tr \ } P(K)=1,
\end{equation}
\begin{equation}
\sum_K P(K)= a_0 1_0\, ,
\end{equation}
where $1_0$ is the projection operator onto
 the sector with total spin $0$. The `completness relation' (2.6) is proven
in appendix A. If we take the trace of eq.(2.6) we get for the
constant $a_0$:
\begin{equation}
a_0=\frac{V!!}{n_0(V)},
\end{equation}
where $V!!$ is the number of valence bond configurations and
$n_0(v)$ the number of spin $0$ states on a lattice with $V=2v$
sites. In appendix A it is shown that spin-spin correlators
$(x,y)$ as they are defined in eq.(2.1) have a particular simple
expansion in terms of the VBPO's:
\begin{equation}
1_0 (x,y) 1_0 =\frac{V+2}{4a_0}\sum_K \delta([x,y]\in K) P(K).
\end{equation}
According to eq.(2.8) all configurations $K$ with a fixed valence
bond $[x,y]$ contribute with equal weight. The remainig configurations
have zero weight. From eq.(2.8) we obtain the expansion of the
nearest neighbour Heisenberg Hamiltonian projected onto the
spin $0$ sector:
\begin{equation}
H_0=1_0\sum_{<x,y>} 4(x,y) 1_0,
\end{equation}
\begin{equation}
H_0=\frac{V+2}{a} \sum_K N_1(K)P(K),
\end{equation}
where
\begin{equation}
N_1(K)=\sum_{<x,y>} \delta([x,y]\in K)
\end{equation}
is the number of dimers -these are the nearest neighbour valence bonds-
on the configuration K. Eq.(2.10) looks like a spectral representation.
But it is not, since the valence bond states $|K\rangle$ are not orthogonal.
The Hamiltonian $H_0$ obeys anticommutation relations with the VBPO's
which are proven in appendix A as well:
\begin{equation}
\frac{1}{2}[H_0,P(K)]_+=(dV+3N_1(K))P(K)+
            \sum_{<x,y>}(P(Q_+(x,y)K)-P(Q_-(x,y)K)).
\end{equation}
The sum on the right hand side extends over configurations which are
obtained from the configuration $K=[x,y_1][x_2,y]...$ by the following
permutations:
\begin{equation}
Q_+(x,y)[x,y_1][x_2,y]=[x,y][x_2,y_1],
\end{equation}
\begin{equation}
Q_-(x,y)[x,y_1][x_2,y]=[y,y_1][x_2,x].
\end{equation}
Eqs.(2.5), (2.6) and (2.12) will provide us with an evolution equation
of the Boltzmann factor between valence bond states.
\section{Projection Operators in Spin $S$ Sectors}
For the construction of states with total spin $S$, we follow
the original work of Hulth\'en \cite{hul}:
\begin{equation}
|K_S\rangle=[z_1,z_2,...z_{2S}]_s [x_1,y_1]...[x_v,y_v].
\end{equation}
Here $[z_1,z_2,...z_{2S}]_s$ denotes a totally symmetrized state
of spins at sites $z_1,...z_{2S}$. As before the $[x_j,y_j]$
are valence bond (spin $0$) states. Therefore we can characterize
each of these states by a configuration $K_S$ with $v=V/2-S$
valence bonds and $2S$ 'monomers'. Each site of the lattice is
occupied either by one valence bond or one monomer.
The corresponding projection operators
\begin{equation}
P(K_S)=|K_S\rangle\langle K_S|,
\end{equation}
have the following properties:
\begin{equation}
{\rm tr \ } P(K_S)=1,
\end{equation}
\begin{equation}
\sum_{K_S} P(K_S)= a_S 1_S,
\end{equation}
where $1_S$ is the projection operator onto
 the sector with total spin $S$ and:
\begin{equation}
a_S=\frac{\nu_S(V)}{n_S(V)}.
\end{equation}
$\nu_S(V)$ is the number of monomer-valence bond configurations
$K_S$ and
$n_S(V)$ the number of spin $S$ states on a lattice with $V$
sites.
The nearest neighbour Heisenberg Hamiltonian projected onto the
spin $S$ sector
\begin{equation}
H_S=1_S\sum_{<x,y>} 4(x,y) 1_S,
\end{equation}
can be expanded in terms of the projection operators $P(K_S)$:
\begin{equation}
H_S=\frac{V+2}{a_S} \sum_{K_S} N_1(K_S)P(K_S),
\end{equation}
where
\begin{equation}
N_1(K_S)=\sum_{<x,y>} \delta([x,y]\in K_S)
\end{equation}
is the number of dimers on the configuration $K_S$.
The Hamiltonian $H_S$ obeys anticommutation relations with the projection
operators $P(K_S)$,
which are proven in appendix A as well:
\begin{eqnarray}
\frac{1}{2}[H_S,P(K_S)]_+=&&(dV+3N_1(K_S)-4N_2(K_S))P(K_S) \nonumber \\&&+
            \sum_{<x,y>}(P(Q_+(x,y)K_S)-P(Q_-(x,y)K_S)).
\end{eqnarray}
$N_2(K_S)$ is the number of monomer pairs on neighbouring sites $<x,y>$.
The action of the permutation operators $Q_\pm{\sigma}(x,y)$
on the valence bond states is given in eqs.(2.13-14) and depicted
in Fig.1a).
If monomers are
involved, one has the following rules:
\begin{equation}
Q_+(x,z)[z,z_2,...]_s[x,y]=[y,z_2,...]_s[x,z]
\end{equation}
\begin{equation}
Q_-(x,z)[z,z_2,...]_s[x,y]=[x,z_2,...]_s[z,y],
\end{equation}
which are visualized in Fig.1b).
The sum on the right hand side of eq. (3.9)
does not include nearest neigbours
$<x,y>$, which are occupied by dimers or monomer pairs.
Finally there exists the analogue of eq.(2.8) for the expansion
of the spin-spin correlators projected on the spin $S$ sector:
\begin{equation}
1_S (x,y) 1_S =\frac{V+2}{4a_S}\sum_{K_S} \delta([x,y]\in K_S) P(K_S).
\end{equation}
\section{Expectation Values of the Boltzmann Factor between
         Monomer-Valence Bond States}
Let us introduce the following notation for the eigenvalues $h_S$
and eigenstates $|h_S\rangle$ of $H_S$:
\begin{equation}
H_S|h_S\rangle=h_SV|h_S\rangle.
\end{equation}
Furthermore let us consider the expectation values of
the Boltzmann factor $\exp(\beta H_S)$
between monomer-valence bond states $|K_S\rangle$:
\begin{equation}
f(\beta,K_S)=\langle K_S|\exp(\beta H_S)|K_S\rangle
=\sum_{h_S} \exp(\beta Vh_S)\langle h_S|P(K_S)|h_S\rangle .
\end{equation}
Then eq.(3.9) -evaluated between the eigenstates $|h_S\rangle$ -
yields a system of linear differential equations describing the
`evolution' of the expectation values $f(\beta,K_S)$:
\begin{eqnarray}
\frac{\partial}{\partial \beta}f(\beta,K_S)
=&&(dV+3N_1(K_S)-N_2(K_S))f(\beta,K_S) \nonumber \\ && +
    \sum_{<x,y>}(f(\beta,Q_+(x,y)K_S)-f(\beta,Q_-(x,y)K_S)),
\end{eqnarray}
with the initial condition:
\begin{equation}
f(\beta=0,K_S)=1 ,\quad {\rm for\quad all} \quad K_S.
\end{equation}
The expectation values $f(\beta,K_S)$ contain the whole information
on the thermodynamical
and magnetic properties of
the system:
\begin{equation}
{\rm tr \ }(\exp{\beta H_S})=
\frac{1}{a_S}\sum_{K_S} f(\beta,K_S).
\end{equation}
Eq.(4.5) results from the `completeness' relation (3.4) for the
projection operators $P(K_S)$.\\
Similarly one gets for the spin-spin correlators $(x,y)$, weighted with the
Boltzmann factor:
\begin{equation}
{\rm tr \ }[(x,y)\exp\beta H_S]=
\frac{V+2}{4a_S}\sum_{K_S} \delta([x,y]\in K_S)f(\beta,K_S)
\end{equation}
as can be derived from the representation (3.12).\\
In the combined limit:
\begin{equation}
V,S \rightarrow \infty , \quad M=S/V, \quad \beta \quad {\rm fixed}
\end{equation}
the free energy per site at fixed inverse temperature $\beta$
and magnetization $M$ is given by:
\begin{equation}
F(\beta,M)=\frac{1}{\beta V} \log {\rm tr \ }(\exp\beta H_S).
\end{equation}
Note that in the absence of an external magnetic field the magnetization
$M$ (in the antiferromagnetic system) is zero. This means that the
free energy  $F(\beta,M=\frac{S}{V}=0)$ in the spin $S=0$ sector contains
already the whole information on the thermodynamics in the absence
of an external field. Of course the same holds for the free energies
in any of the sectors with fixed total spin $S$ and for consistency
they have to be the same. This can be seen immediately from the
evolution equations (4.3), which differ for the various values of $S$
by the number of monomers in the configurations $K_S$. It is at least
plausible that a finite number of monomers will not change the
behavior of the $f(\beta,K_S)$ in the thermodynamical limit.

\section{The Monomer Dimer Cluster Expansion}
In this section, we want to investigate an approximation scheme
for the evolution equation (4.3).
Let us first ask, what are the most important
characteristics of the monomer valence bond configuration $K_S$ which
determine
the variation of $f(\beta,K_S)$ with $K_S$. Obviously, such  characteristic
variables are the densities of dimers and monomer pairs $x_1=N_1/V$,
$x_2=N_2/V$ which appear explicitly
in the evolution equation (4.3). It is clear from eq.(4.3) that
the $f(\beta,K_S)$ increase much stronger with $\beta$
on configurations with high dimer- and low monomer pair
density. Next the clustering of
the monomers and dimers seems to be important as will be demonstrated below.
Therefore, the expectation values
$f(\beta,K_S)$ are supposed to depend on the numbers
$N_1(K_S)$, $N_2(K_S),...$
of dimers, monomer pairs,
neighbouring dimer pairs, etc. \\
Let us discuss first the lowest order in this approximative scheme, where
we average the evolution equation (4.3) over all monomer
valence bond configurations
with fixed numbers $N_1(K_S)$, $N_2(K_S)$ of dimers and monomer pairs.
The approximation would be exact
if the right hand side could be expressed as well in terms of the averages
\begin{equation}\label{averages}
f(\beta,N_1,N_2)=\frac{1}{\nu(N_1,N_2,V)}\sum_{K_S(N_1,N_2)}
f(\beta,K_S(N_1,N_2)).
\end{equation}
This is not possible for all the terms in the sum on the right
hand side of eq.(4.3). Here we approximate the $f(\beta,K_S)$ by their average
(\ref{averages}). The approximate evolution equation can be brought in the
form:
\begin{eqnarray}\label{ee}
\frac{\partial}{\partial \beta}f(\beta,N_1,N_2)
=&&(dV+3N_1-N_2)f(\beta,N_1,N_2)\nonumber \\ && +
V\sum_{j_1,j_2} c_{j_1,j_2}(N_1,N_2,V) f(\beta,N_1+j_1,N_2+j_2).
\end{eqnarray}
The coefficients $c_{j_1,j_2}(N_1,N_2,V)$ are given by :
\begin{eqnarray}\label{cjj}
c_{j_1,j_2}(N_1,N_2,V)=&&\frac{1}{V\nu(N_1,N_2,V)}
 \sum_{<x,y>}\sum_{K(N_1,N_2)}\sum_{\sigma=+,-} \\ &&\sigma \prod_{l=1,2}
 \delta(N_l(K)+j_l-N_l(Q_{\sigma}(x,y)K)). \nonumber
\end{eqnarray}
Note that the permutations $Q_{\pm}(x,y)$ (eqs.(2.13,14),(3.10,11))
-applied
on the monomer valence bond configuration $K_S(N_1)$ -
can change the number of
dimers and monomer pairs:
\begin{equation}
N_1(Q_{\pm}(x,y)K_S)-N_1(K_S)=j_{1\pm}(K_S)=-2,-1,0,1,2
\end{equation}
\begin{equation}
N_2(Q_{\pm}(x,y)K_S)-N_2(K_S)=j_{2\pm}(K_S)=-(d-1),...,(d-1).
\end{equation}
The coefficients (5.3) are investigated in appendix B. They can be expressed
in terms of probabilities to find certain clusters on the monomer valence bond
configurations $K_S(N_1,N_2)$. In the combined limit:
\begin{equation}\label{limit}
V,S,N_1,N_2 \rightarrow \infty , \quad x_1=N_1/V, \quad x_2=N_2/V,
\quad \beta, \quad M=S/V \quad {\rm fixed},
\end{equation}
they only depend on the densities $x_1,x_2$ of dimers and monomerpairs:
\begin{equation}
 c_{j_1j_2}(N_1,N_2,V) \to c_{j_1j_2}(x_1,x_2).
\end{equation}
Note, that the magnetization $M=S/V$ - i.e. the monomer densities - is held
fixed.
It is a conserved quantity in the evolution process.\\
The coefficients (\ref{cjj}) acquire a particularly simple form in the limit:
\begin{equation}
 M \to 0, \quad x_2 \to 0, \quad c_{j_1j_2}(x_1,x_2) \to c_{j_1}(x_1)
\delta_{j_2 0},
\end{equation}
where
\begin{eqnarray}
c_{-2}(x_1)  &=& -w_1(x_1), \\
c_{-1}(x_1)  &=& 3w_1(x_1)- w_2(x_1) - 2(2d-1)x_1,  \\
c_{0}(x_1)   &=& -3w_1(x_1)+w_2(x_1)+4(2d-1)x_1 - d + x_1, \\
c_{1}(x_1) &=& w_1(x_1)+d-x_1(2(2d-1)+1).
\end{eqnarray}
Here $w_1(x),w_2(x)$ are the probabilities to find on the valence bond
configuration
$K_S$:
\begin{itemize}
\item dimer pairs with arbitrary orientation: $w_1(x_1)$
\item dimer pairs with parallel orientation: $w_2(x_1)$.
\end{itemize}
These types of clusters are shown in Fig. 2a-b for the case $d=2$.
\section{The Evolution Equation in the Thermodynamical Limit}
In this section we are going to study the approximate
evolution equation (\ref{ee}) in the combined limit (\ref{limit}).
Then (\ref{ee}) reduces
to  a partial differential equation for
if we assume that the averages (\ref{averages}) have the form:
\begin{equation}
f(\beta,N_1,N_2) = \exp (V \phi(\beta,x_1,x_2)),
\end{equation}
\begin{equation}
\frac{\partial}{\partial x_l}\phi(\beta,x_1,x_2)=\log R_l \quad l=1,2,
\end{equation}
\begin{equation}
\frac{\partial}{\partial \beta}\phi(\beta,x_1,x_2)= L(x_1,R_1,x_2,R_2),
\end{equation}
where
\begin{equation}
L(x_1,R_1,x_2,R_2)=d+3x_1 - x_2+ \sum_{j_1,j_2} c_{j_1,j_2}(x_1,x_2)
                  R_1^{j_1}R_2^{j_2}.
\end{equation}
The initial condition:
\begin{equation}
\phi(\beta=0,x_1,x_2)=0
\end{equation}
is a consequence of eq.(4.4).
To solve eqs.(6.2-5) we first introduce -via a Legendre transform-
a function:
\begin{equation}
\Phi(x_1,R_1,x_2,R_2)=\phi(\beta,x_1,x_2)
-\beta\frac{\partial \phi(\beta,x_1,x_2)}{\partial \beta},
\end{equation}
which depends on the four variables $x_1,x_2,R_1,R_2$. The partial
derivatives of $\Phi$ with respect to these variables are not independent
but obey the following four relations:
\begin{equation}
\beta=-\frac{\partial\Phi}{\partial R_l}
     \left(\frac{\partial L}{\partial R_l} \right)^{-1} \quad l=1,2,
\end{equation}
\begin{equation}
\beta=-\left(\frac{\partial\Phi}{\partial x_l}-\log R_l\right)
     \left(\frac{\partial L}{\partial x_l} \right)^{-1} \quad l=1,2.
\end{equation}
A solution of these equations is found with the ansatz:
\begin{eqnarray}
\Phi(x_1,R_1,x_2,R_2)&&=x_1\log R_1+x_2\log R_2
-\int\limits_{1}^{R_1} \int\limits_{1}^{R_2}
       \frac{dR'_1 dR'_2}{R'_1R'_2} g(R'_1,R'_2,L) \nonumber \\ &&
-\int\limits_{1}^{R_1} \frac{dR'_1}{R'_1} g_1(R'_1,L)
-\int\limits_{1}^{R_2} \frac{dR'_2}{R'_2} g_2(R'_2,L),
\end{eqnarray}
if the integrands $g(R'_1,R'_2,L),g_1(R'_1,L),g_2(R'_2,L)$ with
$L=L(x_1,R_1,x_2,R_2)$ satisfy
the conditions:
\begin{equation}
x_1=x_1(R_1,R_2,L)=\int\limits_{1}^{R_2}
\frac{dR'_2}{R'_2}g(R_1,R'_2,L)+g_1(R_1,L),
\end{equation}
\begin{equation}
x_2=x_2(R_1,R_2,L)=\int\limits_{1}^{R_1}
\frac{dR'_1}{R'_1}g(R'_1,R_2,L)+g_2(R_2,L).
\end{equation}
Note that the ansatz (6.9) satisfies the initial condition (6.5).
According to (6.2), the initial point $\beta=0$ is mapped on the point
$R_1=R_2=1$ in the $R_1,R_2$-plane. There the Legendre transform (6.6)
has to vanish for all $x_1,x_2$.\\
Eqs.(6.10,11) define $x_1,x_2$ as fuctions of $R_1,R_2,L$ with partial
derivatives:
\begin{equation}
R_2\left.\frac{\partial x_1}{\partial R_2}\right|_{R_1,L}
=R_1\left.\frac{\partial x_2}{\partial R_1}\right|_{R_2,L}=g(R_1,R_2,L).
\end{equation}
Therefore $x_l,l=1,2$ can be written as the gradient:
\begin{equation}
x_l=\frac{\partial}{\partial \log R_l}G(R_1,R_2,L)
\end{equation}
of a `potential' $G(R_1,R_2,L)$. Inserting (6.13) into (6.4) we end
up with a partial differential equation for $G(R_1,R_2,L)$. The
solution of this equation demands the knowledge of the coefficients
$c_{j_1,j_2}(x_1,x_2)$ in (6.4). They will be computed in the
subsequent papers, where we treat the thermodynamical and magnetic
properties of the antiferromagnetic Heisenberg model in dimensions
$d=1,2,3$.
\section{Summary and Perspectives}
The main result of this paper is the evolution equation (4.3) for the
expectation values (4.2) of the Boltzmann factor between
monomer valence bond states. Once the evolution equation is solved,
thermodynamical and magnetic properties (cf.(4.5)) as well as
spin-spin correlators (cf.(4.6)) can be easily computed.
The high temperature expansion of eq.(4.3) yields a  recursion relation
in a vector space of dimension $\nu_S(V)$.
This is just the number of monomer valence bond configurations.
This number can
be reduced by using the symmetries of the Hamiltonian $H_S$.
It seems to be realistic that a direct solution of such a recursion relation
is feasible on clusters up to 18 sites.\\
On the other hand, the exact evolution equation (4.3) can be considered
as the starting point of a new type of approximation for the quantum
Heisenberg model. In the thermodynamical limit the problem of
solving the approximated eqation (5.2) can be reduced to
the solution of a partial differential equation. The approximation
of the evolution equation (4.3) can be systematically improved. The
prize one has to pay is that the corresponding partial differential
equations become more complex.\\
We meet the most simple situation in the spin $0$ sector, where the
approximated evolution equation (5.2) can be solved analytically-
up to integrations. We will discuss this solution in a second paper
and derive from it explicit expressions for the specific heat
of the antiferromagnetic Heisenberg model in dimensions $d=1,2,3$.

\newpage

\begin{appendix}
\section{The Properties of the Valence Bond Projection Operators (VBPO)}
The properties of the VBPO's listed in eqs.(2.6), (2.8), (2.10) and
(2.12) are based on the following anticommutation relations for the
projection operators $(x,y)$ defined in eq.(2.1):
\begin{equation}
(x,y)^2=(x,y)
\end{equation}
\begin{equation}
\frac{1}{2}[(1,2),(2,3)]_+=\frac{1}{4}((1,2)+(2,3)-(1,3))
\end{equation}
\begin{equation}
\frac{1}{2}[(1,2)(3,4),(2,3)]_+=\frac{1}{4}
((1,2)(3,4)+(1,4)(2,3)-(1,3)(2,4)).
\end{equation}
For the proof of the completeness relation (2.6) it is sufficient
to show that the sum over all VBPO's:
\begin{equation}
\sum_K P(K)=\sum_j a_j(\vec{S}^2)^j
\end{equation}
can be written as a power series in the square of the total spin $\vec{S}$.
Eq.(A4) obviously implies eq.(2.6): The right hand side -applied to
any spin $0$ state- produces the same state. The left hand side -applied
to any state with spin $s>0$- yields zero. To verify (A4) we start with
the sum:
\begin{equation}
\Sigma_j=\sum_{x_1,...,y_j}(x_1,y_1)(x_2,y_2)...(x_j,y_j)
\end{equation}
over the product of $j$ disconnected valence bonds. We are going to
show by complete induction that:
\begin{equation}
\Sigma_j=\sum_{l=0}^{l=j} a_{jl} (\vec{S}^2)^l
\end{equation}
has the property we are looking for. Eq.(A6) is obviously correct for
$j=1$:
\begin{equation}
\Sigma_1=\sum_{x\neq y}(x,y)=a_{10}+a_{11}\vec{S}^2.
\end{equation}
According to the definition (A5), the anticommutator
$[\Sigma_j,\Sigma_1]_+$ contains for types of contributions:
\begin{itemize}
\item disconnected: The sites $x,y$ do not overlap with the sites
 $x_1....y_j$. The sum over these contributions yields $\Sigma_{j+1}$.
\item one site connected: One of the sites $x$ or $y$ overlaps with
the sites $x_1...y_j$. The sum over these contributions leads back
to terms of the type $\Sigma_j$, as can be verified with the
anticommutator (A2).
\item two site connected: Both sites $x$ and $y$ overlap with
$x_1,...,y_j$. Again the sum over these contributions leads back to
terms of the type $\Sigma_j$ as can be seen from the anticommutator
(A3).
\end{itemize}
This completes the proof of eq.(A4).\\
Next we turn to the proof of the representation (2.8). From eqs.(A1)
and (A3) we get anticommutation relations for the VBPO's on
two types of configurations:
\begin{itemize}
\item $K_1$: sites $x$ and $y$ are occupied by the same valence bond
\begin{equation}
\frac{1}{2}[P(K_1),(x,y)]_+=P(K_1)
\end{equation}
\item $K_2$: sites $x$ and $y$ are occupied by two different valence bonds
\begin{equation}
\frac{1}{2}[P(K_2),(x,y)]_+=\frac{1}{4}(P(K_2)+P(Q_+(x,y)K_2)-P(Q_-(x,y)K_2)).
\end{equation}
\end{itemize}
The action of the permutations $Q_\pm{\sigma}(x,y),$ is defined
in eqs.(2.13,14). From eq.(A9) we get
\begin{equation}
[P(K_2)+P(Q_-(x,y)K_2),(x,y)]_+= P(Q_+(x,y)K_2).
\end{equation}
and by summation over all configurations of the type $K_2$
\begin{equation}
\sum_K[P(K),(x,y)]_+-\sum_{K_1}[P(K_1),(x,y)]_+ =
    \frac{V-2}{2}\sum_{K_1}P(K_1).
\end{equation}
On the right hand side we have used the fact that the configuration
$Q_+(x,y)K_2$ is of the type $K_1$. Evaluation of the left hand side
by means of the completeness relation (2.6) and the projector
property (A8) leads to the representation (2.8). Finally, the
anticommutation relations (2.12) for the Hamiltonian $H_0$ with the
VBPO's are derived from eqs.(A8) and (A9) by summation over all
nearest neighbour valence bonds $(x,y)$.\\
The properties (3.4), (3.9) and (3.12) of the projection operators
$P(K_S)$ can be derived in a similar way. For the proof of the
completeness relation (3.4) we express the projection operators:
\begin{equation}
|[z_1,...,z_{2S}]_+ \rangle\langle [z_1,...z_{2S}]_+|=c\prod_{j=1}^{S-1}
             (\vec{s}^{\ 2}-j(j+1))
\end{equation}
as a power series of the square of the total spin $\vec{s}$:
\begin{equation}
\vec{s}=\sum_{l=1}^{2S} \vec{s}(z_l)
\end{equation}
on the sublattice which is occupied by the monomers at sites
$z_1,...z_{2S}$. From the decomposition (A6) we know, that any
power of $\vec{s}^{\ 2}$:
\begin{equation}
(\vec{s}^{\ 2})^j=\sum_{l=0}^j c_{jl} \Sigma'_l
\end{equation}
can be written as a sum over terms $\Sigma'_l$ of the type (A5).
In $\Sigma'_l$ we sum over the product of $l$ valence bonds on
the sublattice with sites $z_1,...z_{2S}$.\\
The sum over all projection operators $P(K_S)$:
\begin{equation}
\sum_{K_S} P(K_S)=\sum_{j=0}^{\frac{V}{2}}b_j \Sigma_j
                 =\sum_{j=0}^{\frac{V}{2}}c_j (\vec{S}^2)^j
\end{equation}
is then expressed in terms of the quantities $\Sigma_j$ defined
on the whole lattice. Eq.(A5) yields the expansion in powers
of $\vec{S}^2$, which is sufficient for the proof of the
completeness relation (3.4).\\
For the proof of the anticommutation relation (3.9) and the
representation (3.12) we have to generalize eqs.(A8) and (A9)
to the corresponding equations for the projection operators
$P(K_S)$. Here, we have to consider three types of configurations:
\begin{itemize}
\item $K_1$: sites $x$ and $y$ are occupied by the same valence bond
\begin{equation}
\frac{1}{2}[P(K_1),(x,y)]_+=P(K_1).
\end{equation}
\item $K_2$: sites $x$ and $y$ are occupied either by two different
valence bonds or by one valence bond and one monomer
\begin{equation}
\frac{1}{2}[P(K_2),(x,y)]_+=\frac{1}{4}(P(K_2)+P(Q_+(x,y)K_2)-P(Q_-(x,y)K_2).
\end{equation}
The action of the permutations $Q_{\pm}(x,y)$ is defined
for the first case in (2.13,14) and for the second one in (3.10,11).
\item $K_3$: sites $x$ and $y$ are occupied by two monomers
\begin{equation}
\frac{1}{2}[P(K_3),(x,y)]_+=0 .
\end{equation}
\end{itemize}
{}From eq.(A17) we get
\begin{equation}
[P(K_2)+P(Q_-(x,y)K_2),(x,y)]_+=P(Q_+(x,y)K_2),
\end{equation}
and by summation over all configurations $K_2$
\begin{eqnarray}
\sum_{K_S}[P(K_S),(x,y)]_+&&-\sum_{K_1}[P(K_1),(x,y)]_+
-\sum_{K_3}[P(K_3),(x,y)]_+
\\ &&=\frac{V-2}{4}\sum_{K_1} P(K_1). \nonumber
\end{eqnarray}
Finally, the representation (3.12) and the anticommutation relation (3.9)
follow
from (A16), (A18), (A20)
and the completeness relation (3.4).
\section{Monomer Dimer Clusters on the configurations $K_S$}
We are going now to exploit equation (5.3) for the coefficients in the
approximated evolution equation (\ref{ee}).  The first sum on the
right hand side of eq.(5.3) extends over those nearest neighbours
$<x,y>$ which are not occupied by a dimer or a monomer pair. The
following situations -- which are depicted for $d=2$ and $<x,y>=<2,3>$
in Fig. 2a-f)-are left:
\begin{itemize}
\item $K_a$: dimer-dimer ( not parallel)
\item $K_b$: dimer-dimer (parallel)
\item $K_c$: dimer-valence bond
\item $K_d$: valence bond - valence bond
\item $K_e$: dimer-monomer
\item $K_f$: valence bond-monomer
\end{itemize}
The valence bonds in the situations $K_c,K_d,K_f$ must not be dimers
in order to avoid double counting.
On the configurations $K_a,K_b,K_c,K_d$ the neighbouring sites
$<x,y> = <2,3>$ are occupied only by dimers and valence bonds. Therefore
the permutations $Q_{\pm}(2,3)$ will not move the  monomers on the
configurations, which means that the number of monomer pairs does not change:
\begin{equation}
N_2(Q_{\pm}(2,3) K_l) - N_2(K_l) = 0 \quad\text{for}\quad l=a,b,c,d.
\end{equation}
The changes of the number of dimers (eq.(5.4)) on these configurations can
be read of table I.

\begin{table}
\caption{}
\begin{tabular}{ccccccc}
&$K_a$& $K_b$& $K_c$& $K_d$&
             $K_e(\mu(1),\mu(2),\mu(3))$&$K_f(\mu(x),\mu(2),\mu(3))$\\ \hline
$j_{1+}$& -1&  0&  0&  1&  0&0  \\
$j_{1-}$& -2& -2& -1&  0& -1&0 \\
$j_{2+}$&  0&  0&  0& -1&  $\mu(1)-\mu(3)$   & $\mu(x)-\mu(3)$   \\
$j_{2-}$&  0&  0&  0& -1&  $\mu(2)-\mu(3)-1$ & $\mu(2)-\mu(3)-1$  \\
\end{tabular}
\end{table}

Let us now turn to the configurations $K_e$ where monomers are involved. The
permutations $Q_+(2,3)$ and $Q_-(2,3)$ will move the monomers from site 3 to
site 1 and site 2, respectively -- according to the rules given in Fig. 1b.
The resulting change (5.5) in the number of monomer pairs depends on the
monomers surrounding the sites 1, 2, and 3 on the configuration
$K_e = K_e(\mu(1),\mu(2),\mu(3))$. If we denote by $\mu(x), x=1,2,3$ the
number of monomers on the neighbouring sites of $x$ we find:
\begin{eqnarray}
N_2(Q_+(2,3)K_e) - N_2(K_e) &=& \mu(1)-\mu(3), \nonumber \\
N_2(Q_-(2,3)K_e) - N_2(K_e) &=& \mu(2)-\mu(3) -1.
\end{eqnarray}
In the same way, one can see that the numbers $\mu(x),\mu(2),\mu(3)$ on the
neighbouring sites of $ x,2,3$ determine the change in the number of monomer
pairs on the configuration $K_f = K_f(\mu(x),\mu(2),\mu(3))$. \\
The changes (5.4,5.5) in the numbers of dimers and monomer pairs -- due to
the action of the permutations $Q_+(2,3)$ and $Q_-(2,3)$ on the configuration
$K_l\quad l=a,...,f$ -- is summarized in table I.\\
Next we need the probabilities $w_l \quad (l=a,...,f)$ to find these
situations $K_l$ on the monomer valence bond configurations $K_S(N_1,N_2)$.
In principal, they can be determined by a Monte Carlo simulation of the
monomer valence bond system. The situation simplifies considerably in the
spin 0 sector, where we have no monomers and where the complicated situations
 $K_e$ and $K_f$ do not occur. The remaining situations $K_a - K_d$ have
probabilities:
\begin{eqnarray}
 w_a &=&  w_1 - w_2, \\
 w_b &=&  w_2, \\
 w_c &=&  2(2d-1)x_1-2w_1, \\
 w_d &=&  (d-x_1)-w_c-w_1, \\
\end{eqnarray}
which can be completely expressed in terms of the dimer pair probabilities
$w_1(x_1),w_2(x_2)$ defined in section V. The coefficients (5.9-12) then
follow from
table I:
\begin{eqnarray}
 c_{-2} &=& -w_a - w_b, \\
 c_{-1} &=&  w_a - w_c, \\
 c_{0}  &=&  w_b + w_c - w_d, \\
 c_{1}  &=&  w_d. \\
\end{eqnarray}
\end{appendix}

{\bf FIGURE CAPTIONS}
\begin{enumerate}
\item The action of the permutations $Q_{\pm}(x,y) $ on
a monomer-valence bond configuration $K_S$:
\begin{itemize}
\item a) both sites $x$ and $y$ are occupied by valence bonds
\item b) site $z$ is occupied by a monomer, site $x$ by a valence bond
\end{itemize}
\item Different types of monomer-valence bond clusters at the nearest
neighbour points $2$ and $3$.
\end{enumerate}
\newpage
\renewcommand{\baselinestretch}{1.0}
\setlength{\baselineskip}{12pt}
\small\normalsize
\textheight=25cm
\textwidth=15.4cm
\topmargin 0.0cm
\headsep 1.0cm
\voffset=-0.5cm
\oddsidemargin=0.0cm

$K_s : $\vspace{-0.5cm}

\unitlength1.0mm
\begin{picture}(150,-6)
\put(25,0){\line(1,0){30}}
\put(25,2){$y_1$}
\put(54,2){$x$}
\put(70,0){\line(1,0){30}}
\put(70,2){$y$}
\put(99,2){$x_2$}
\end{picture}
\vspace{1cm}

$Q_+(x,y) K_s : $\vspace{-0.8cm}

\unitlength1.0mm
\begin{picture}(150,6)
\put(25,0){\line(1,0){75}}
\put(25,2){$y_1$}
\put(54,4){$x$}
\put(55,2){\line(1,0){15}}
\put(70,4){$y$}
\put(99,2){$x_2$}
\end{picture}
\vspace{1.6cm}

$Q_-(x,y) K_s : $\vspace{-0.8cm}

\unitlength1.0mm
\begin{picture}(150,6)
\put(25,0){\line(1,0){45}}
\put(25,-3){$y_1$}
\put(54,4){$x$}
\put(55,2){\line(1,0){45}}
\put(70,-3){$y$}
\put(99,4){$x_2$}
\end{picture}
\vspace{1cm}

\centerline{\large Fig. 1a)}

%\end{figure}

\vspace{2cm}

%\begin{figure}[h]

$K_s :$\vspace{-0.5cm}

\unitlength1.0mm
\begin{picture}(150,6)
\put(30,0){\circle*{1}}
\put(45,0){\line(1,0){30}}
\put(30,2){$z$}
\put(44,2){$x$}
\put(75,2){$y$}
\end{picture}
\vspace{1.2cm}

$Q_+(x,z) K_s :$\vspace{-0.8cm}

\unitlength1.0mm
\begin{picture}(150,6)
\put(75,0){\circle*{1}}
\put(30,0){\line(1,0){15}}
\put(30,2){$z$}
\put(44,2){$x$}
\put(75,2){$y$}
\end{picture}
\vspace{1cm}

$Q_-(x,z) K_s :$\vspace{-0.8cm}

\unitlength1.0mm
\begin{picture}(150,6)
\put(45,2){\circle*{1}}
\put(30,0){\line(1,0){45}}
\put(30,2){$z$}
\put(44,4){$x$}
\put(75,2){$y$}
\end{picture}
\vspace{1cm}

\centerline{\large Fig. 1b)}
\newpage

$K_a : $\vspace{-0.8cm}

\unitlength1.0mm
\begin{picture}(150,6)
\put(9,2){1}
\put(10,0){\line(1,0){10}}
\put(19,2){2}
\multiput(22,0)(2,0){4}{\circle*{0.5}}
\put(29,2){3}
\put(30,0){\line(1,0){10}}
\put(39,2){4}
\put(55,-1){2}
\put(60,0){\line(0,-1){10}}
\put(55,-11){1}
\multiput(62,0)(2,0){4}{\circle*{0.5}}
\put(72,-1){3}
\put(70,0){\line(0,1){10}}
\put(72,9){4}
\put(85,-1){2}
\put(90,0){\line(0,1){10}}
\put(85,9){1}
\multiput(92,0)(2,0){5}{\circle*{0.5}}
\put(102,-1){3}
\put(100,0){\line(0,-1){10}}
\put(102,-11){4}
\end{picture}
\vspace{1.6cm}

\begin{picture}(150,6)
\put(10,0){\line(1,0){10}}
\multiput(22,0)(2,0){4}{\circle*{0.5}}
\put(30,0){\line(0,1){10}}
\put(9,2){1}
\put(19,2){2}
\put(32,0){3}
\put(32,9){4}
\put(50,0){\line(1,0){10}}
\multiput(62,0)(2,0){4}{\circle*{0.5}}
\put(70,0){\line(0,-1){10}}
\put(49,2){1}
\put(59,2){2}
\put(72,0){3}
\put(72,-11){4}
\put(90,0){\line(0,1){10}}
\multiput(92,0)(2,0){4}{\circle*{0.5}}
\put(100,0){\line(1,0){10}}
\put(86,0){2}
\put(86,9){1}
\put(99,2){3}
\put(109,2){4}
\put(130,0){\line(0,-1){10}}
\multiput(132,0)(2,0){4}{\circle*{0.5}}
\put(140,0){\line(1,0){10}}
\put(126,0){2}
\put(126,-11){1}
\put(139,2){3}
\put(149,2){4}
\end{picture}

\vspace{2.7cm}

$K_b : $\vspace{-0.8cm}

\begin{picture}(150,6)
\unitlength1.0mm
\put(20,0){\line(0,1){10}}
\multiput(22,0)(2,0){4}{\circle*{0.5}}
\put(30,0){\line(0,1){10}}
\put(16,9){1}
\put(16,0){2}
\put(32,0){3}
\put(32,9){4}
\put(50,0){\line(0,-1){10}}
\multiput(52,0)(2,0){4}{\circle*{0.5}}
\put(60,0){\line(0,-1){10}}
\put(46,-11){1}
\put(46,0){2}
\put(62,0){3}
\put(62,-11){4}
\end{picture}
\vspace{2.7cm}

$K_c : $\vspace{-0.8cm}

\begin{picture}(150,6)
\put(10,0){\line(1,0){10}}
\multiput(22,0)(2,0){4}{\circle*{0.5}}
\put(30,0){\line(1,0){20}}
\put(9,2){1}
\put(19,2){2}
\put(29,2){3}
\put(49,2){x}
\put(70,0){\line(0,1){10}}
\multiput(72,0)(2,0){4}{\circle*{0.5}}
\put(80,0){\line(1,0){20}}
\put(66,9){1}
\put(66,0){2}
\put(79,2){3}
\put(99,2){x}
\put(120,0){\line(0,-1){10}}
\multiput(122,0)(2,0){4}{\circle*{0.5}}
\put(130,0){\line(1,0){20}}
\put(116,-11){1}
\put(116,0){2}
\put(129,2){3}
\put(149,2){x}
\end{picture}
\vspace{1.6cm}

\begin{picture}(150,6)
\put(10,0){\line(1,0){20}}
\multiput(32,0)(2,0){4}{\circle*{0.5}}
\put(40,0){\line(1,0){10}}
\put(9,2){x}
\put(29,2){2}
\put(39,2){3}
\put(49,2){4}
\put(70,0){\line(1,0){20}}
\multiput(92,0)(2,0){4}{\circle*{0.5}}
\put(100,0){\line(0,1){10}}
\put(69,2){x}
\put(89,2){2}
\put(103,0){3}
\put(103,9){4}
\put(120,0){\line(1,0){20}}
\multiput(142,0)(2,0){4}{\circle*{0.5}}
\put(150,0){\line(0,-1){10}}
\put(119,2){x}
\put(139,2){2}
\put(153,0){3}
\put(153,-11){4}
\end{picture}
\vspace{2.8cm}

$K_d : $\vspace{-0.8cm}

\begin{picture}(150,6)
\put(10,0){\line(1,0){20}}
\multiput(32,0)(2,0){4}{\circle*{0.5}}
\put(40,0){\line(1,0){20}}
\put(9,2){x}
\put(29,2){2}
\put(39,2){3}
\put(59,2){y}
\end{picture}

\vspace{2.8cm}

$K_e : $\vspace{-0.8cm}

\begin{picture}(150,6)
\put(10,0){\circle{1}}
\put(20,10){\circle{1}}
\put(20,-10){\circle{1}}
\put(30,10){\circle{1}}
\put(30,-10){\circle{1}}
\put(40,10){\circle{1}}
\put(40,-10){\circle{1}}
\put(50,0){\circle{1}}
\put(40,0){\circle*{1}}
\put(20,0){\line(1,0){10}}
\multiput(32,0)(2,0){4}{\circle*{0.5}}
\put(19,2){1}
\put(29,2){2}
\put(39,2){3}
\end{picture}
\vspace{2.8cm}

$K_f : $\vspace{-0.8cm}

\begin{picture}(150,6)
\put(10,0){\circle{1}}
\put(20,10){\circle{1}}
\put(20,-10){\circle{1}}
\put(30,0){\circle{1}}
\put(50,0){\circle{1}}
\put(60,10){\circle{1}}
\put(60,-10){\circle{1}}
\put(70,10){\circle{1}}
\put(70,-10){\circle{1}}
\put(80,0){\circle{1}}
\put(70,0){\circle*{1}}
\put(20,0){\line(1,0){40}}
\multiput(62,0)(2,0){4}{\circle*{0.5}}
\put(19,2){x}
\put(59,2){2}
\put(69,2){3}
\end{picture}
\vspace{2.0cm}
\centerline{\large Fig. 2a-f)}

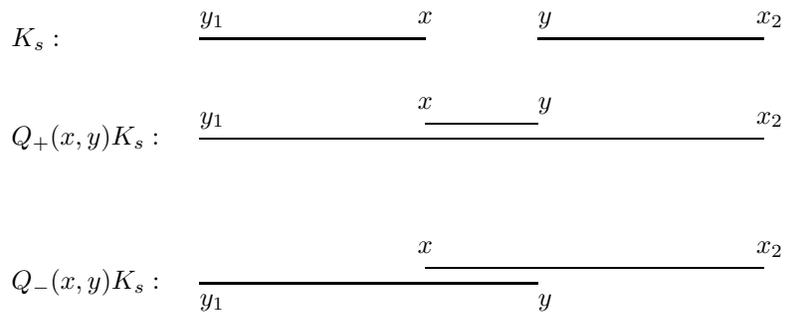
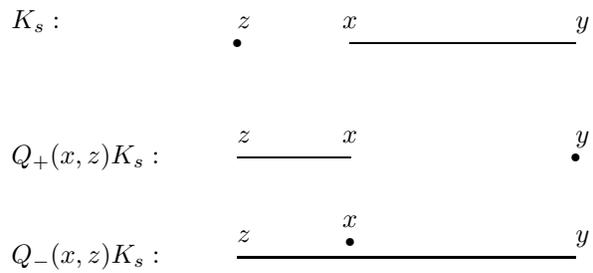
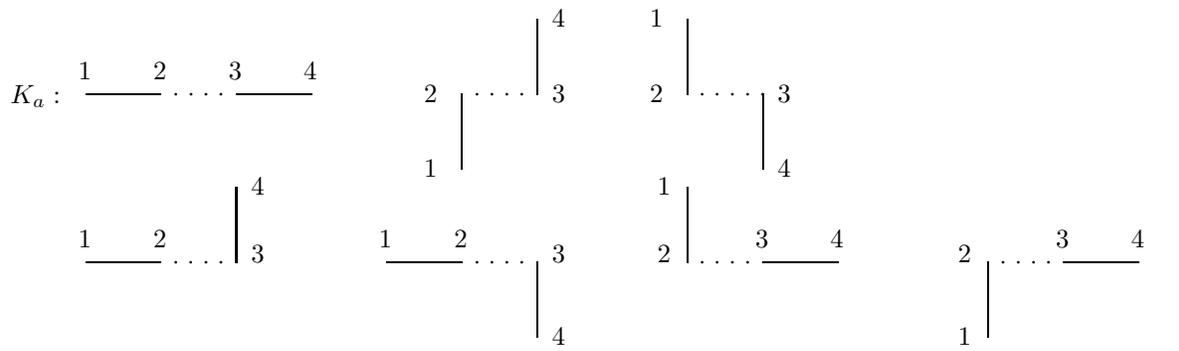
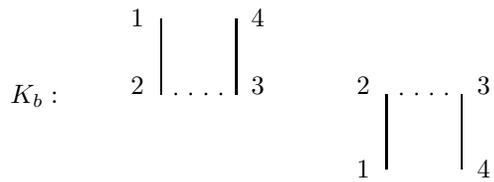
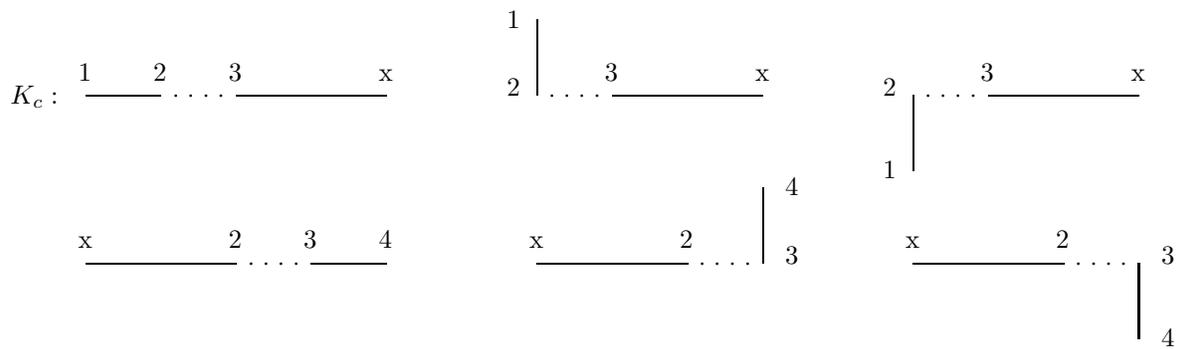
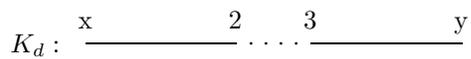
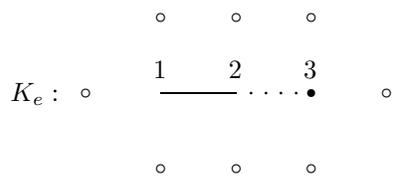
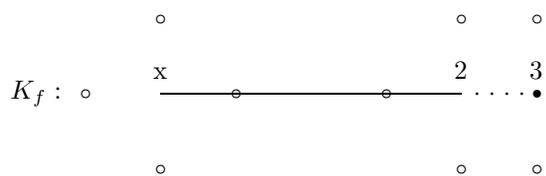
\end{document}